\documentclass[10pt, conference, letterpaper]{IEEEtran}

\makeatletter

\renewcommand{\baselinestretch}{0.955}

\makeatletter
\def\ps@headings{%
\def\@oddhead{\mbox{}\scriptsize\rightmark \hfil }
\def\@evenhead{\scriptsize\thepage \hfil \leftmark\mbox{}}%
\def\@oddfoot{}%
\def\@evenfoot{}}
\makeatother
\usepackage{authblk}
\usepackage{epsfig}
\usepackage{epstopdf}
\usepackage{simplemargins}
\usepackage{amsfonts}
\usepackage{amsmath,bm}
\usepackage{graphicx}
\usepackage{url}
\usepackage{subfig}
\usepackage{booktabs}
\usepackage{multicol}
\usepackage{algorithmic}
\usepackage[ruled,commentsnumbered]{algorithm2e}
\usepackage{booktabs}
\usepackage{amssymb}
\usepackage{amsthm}

\newtheorem{theorem}{Theorem}
\newtheorem{lemma}{Lemma}
\newtheorem{definition}{Definition}

\IEEEoverridecommandlockouts
\usepackage{cite}
\usepackage{algorithmic}
\usepackage{textcomp}
\def\BibTeX{{\rm B\kern-.05em{\sc i\kern-.025em b}\kern-.08em
		T\kern-.1667em\lower.7ex\hbox{E}\kern-.125emX}}

\begin{document}

\title{Learning the Optimal Synchronization Rates in Distributed SDN Control Architectures
\thanks{This publication was supported partly by the U.S. Army Research Laboratory and the U.K. Ministry of Defence under Agreement Number W911NF-16-3-0001 and the Army Research Office under Agreement Number W911NF-18-10-378.}
}

\author[1]{Konstantinos Poularakis}
\author[1]{Qiaofeng Qin}
\author[2]{Liang Ma}
\author[3]{Sastry Kompella}
\author[4]{Kin K. Leung}
\author[1]{Leandros Tassiulas}
\affil[1]{Department of Electrical Engineering and Institute for Network Science, Yale University, USA}
\affil[2]{IBM T. J. Watson Research Center, Yorktown Heights, NY, USA}
\affil[3]{U.S. Naval Research Laboratory, Washington, DC, USA}
\affil[4]{Department of Electrical and Electronic Engineering, Imperial College London, UK}

\maketitle
\begin{abstract}
Since the early development of Software-Defined Network (SDN) technology, researchers have been concerned  with the idea of physical
distribution of the control plane to address scalability and reliability challenges of centralized designs.  However, having multiple controllers managing the network while maintaining a ``logically-centralized" network view brings  additional challenges. One  such challenge is how to coordinate the management decisions made by the controllers which is usually achieved by  disseminating  synchronization messages in a peer-to-peer manner. While there exist many architectures and protocols to ensure synchronized network views and drive coordination among controllers, there is no systematic methodology for deciding the optimal frequency (or rate) of  message dissemination. In this paper, we fill this gap by  introducing  the SDN synchronization problem: how often to synchronize the network views for each controller pair. We consider two different objectives; first, the maximization of the  number of controller pairs that are synchronized, and second, the maximization of the performance of applications of interest which may be affected by the synchronization rate. Using techniques from knapsack optimization and learning theory, we derive algorithms with provable performance guarantees for each objective. Evaluation results demonstrate significant benefits over baseline schemes that synchronize all controller pairs at equal rate.
\end{abstract}

\pagestyle{headings}
\pagenumbering{arabic}
\setcounter{page}{1}
\thispagestyle{empty}

\section{Introduction} \label{section:introduction}

\subsection{Motivation}

Software Defined Networking (SDN) is a rapidly emerging technology that brings new flexibility to network management and therefore facilitates the implementation of advanced traffic engineering mechanisms~\cite{sdn-survey}. The main principle of SDN is to shift all the network control functions from the data forwarding devices to a programmable network entity, the \emph{controller}. To ensure availability in case of controller failure, typical SDN systems deploy multiple controllers. The controllers may be physically distributed across the network, but they should be ``logically-centralized''. This means that the controllers should coordinate their decisions to ensure their collective behavior matches the behavior of a single controller.

The coordination among controllers is an active area of research with several protocols proposed thus far~\cite{distributed-sdn}. For example, OpenDaylight~\cite{odl} and ONOS~\cite{onos}, two state-of-the-art controller implementations, rely on RAFT and Anti-entropy protocols for disseminating coordination messages among controllers. Typically, each controller is responsible for a part of the network only, commonly referred to as the controller's \emph{domain}. The messages disseminated by a controller to the other controllers convey its view on the state of its domain (e.g., available links and installed flows). The composition of these messages  allow the controllers to synchronize and agree on the state of the entire network.

While different coordination protocols may generate messages of different types and at different timescales, there exist two broad protocol categories~\cite{consistency}. The first category contains the \emph{strongly consistent} protocols which strive to maintain all the controllers synchronized in all times. This is ensured by disseminating messages each time a network change (e.g., a node or link failure) happens followed by a consensus procedure. The second category contains the \emph{eventually consistent} protocols which omit the consensus procedure, yet converge to a common state in a timely manner usually through periodic message dissemination.

\begin{figure}[t]
	\begin{center}
		\includegraphics[scale=0.5]{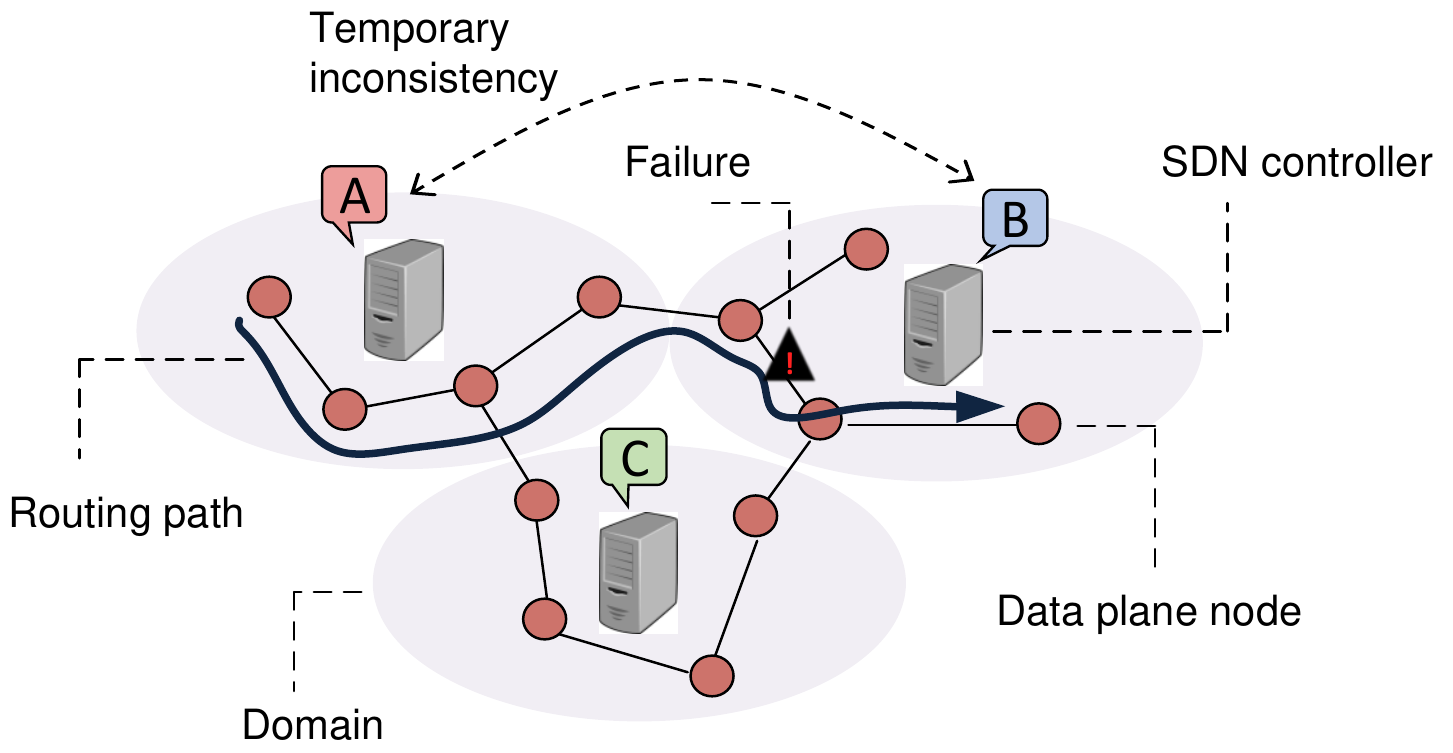}
		\caption{Impact of inconsistency among controllers on routing application performance.}
		\label{fig:example}
	\end{center}
\vspace{-2mm}
\end{figure}

Despite its benefits, strong consistency is difficult to ensure in practice as it is challenged by the unreliable nature of network communications. In addition, this approach generates significant \emph{overheads} for message dissemination among controllers which may be prohibitively large especially when applied to wireless networks with in-band control channels of limited capacity~\cite{onos-traffic},~\cite{infocom-18},~\cite{sdn-planning}. On the other hand, eventual consistency, where controllers are permitted to temporarily have inconsistent views of each other's state, better suits the needs of the above networks, and, thus, can be used to extend the applicability of distributed controller solutions. Yet, the inconsistent views of controller states can harm the \emph{performance of network applications}.

To illustrate the impact of inconsistency, we consider the toy example with three controllers (A, B and C) and their respective domains in Figure \ref{fig:example}. Each pair of controllers synchronize periodically, e.g., every few seconds. At some time, controller A receives a request for routing a flow to a destination node inside the domain of B. Controller A will respond by computing and setting up a routing path based on its current view on the state (topology, traffic loads) of its domain and the other domains. However, controller A is not aware if the links on the routing path outside of its domain are still available or have failed (e.g., a failed link in domain B in Figure \ref{fig:example}) since the last synchronization period. If failures happened, the packets of the flow will have to wait until the next synchronization period, although there is an alternative directly available routing path through the domain of C. Similar problems, if not more serious, can be identified for more advanced traffic engineering applications where inconsistency hinders the effective load balancing and distribution across multiple paths.

The eventually consistent model raises new technical challenges. In particular, it is important to decide \emph{how often} (at what period or rate) to synchronize each pair of controllers in a given network. One might expect that the straightforward policy where all controller pairs synchronize at the same rate would work well. However, some may argue that the synchronization rate should be higher for domains that are more dynamic (with many changes in topology and flow configurations) in order to preserve consistency of the rest domains.

The issue is further complicated by the requirements of the network applications. Previous works~\cite{levin},~\cite{scl} showed that certain network applications, like load-balancers, can work around eventual consistency and still deliver acceptable (although degraded) performance. In such cases, some additional effort needs to be made to ensure that conflicts such as forwarding loops, black holes and reachability violation are avoided~\cite{guo}. Therefore, synchronization policies that completely neglect the specific applications of interest in the network as well as the impact of synchronization rate on their performance may end-up being highly inefficient.


The above questions remain open since, until now, the inter-controller traffic has been often neglected in SDN literature with most of the existing works focusing on the routing and balancing of the data traffic (e.g., see the survey in \cite{sdn-survey} and the discussion of related work in Section \ref{section:related}).

\subsection{Methodology and Contributions}

Our goal in this paper is to investigate policies for the synchronization among SDN controllers, and focus particularly on the impact of the rate of synchronization  on the performance of network applications.
We begin by introducing a model of a  system with multiple controllers (and domains) that is general enough to capture different synchronization overhead costs, as well as network topologies and domain dynamics. We then utilize this model to derive the optimal synchronization rate policy under a total overhead constraint.  We explicitly consider two objectives. First, we target the maximization of the number of controller pairs that are consistent with each other, referred to as the \emph{consistency level} (Obj. 1). We show that for this objective the synchronization problem is NP-Hard and develop a pseudopolynomial-time optimal as well as a Fully Polynomial Time Approximation (FPTA) algorithm using a connection with the \emph{multiple-choice knapsack (MCK) problem}~\cite{mck}.

The second objective (Obj. 2) aims to maximize the \emph{performance of network applications} rather than the overall consistency level of the system. This is a more complex problem since, in practice, we do not know the function that maps the synchronization rate to application performance. To obtain some quantitative insights on this function, we emulate the performance of two applications of interest, namely shortest path routing and load balancing, using a commercial platform (Mininet)~\cite{mininet} and SDN controller (RYU)~\cite{ryu}. While the results are quite unsteady, the average performance increases with the synchronization rate and saturates eventually showing that a diminishing return rule applies. To overcome the unknown objective challenge, we use elements from the  \emph{learning theory}, and propose an algorithm that gradually trains the system and constructs a solution that is with high confidence close to the optimal.
The contributions of this work can be summarized as follows:

\begin{enumerate}

\item We introduce the problem of finding the optimal synchronization rates among SDN controllers in a network, using a general model and different objectives. To the best of our knowledge, this is the first work that studies this problem.

\item For the consistency level maximization objective (Obj. 1), we show that this problem is NP-Hard and provide a pseudopolynomial-time optimal and a FPTA algorithm.

\item For the application performance maximization objective (Obj. 2),  we emulate the performance of two popular applications and obtain insights about the impact of synchronization rates. We use these results to derive an algorithm that gradually trains the system in order to learn the optimal policy.

\item We perform evaluations to show the efficiency of our proposed algorithms. We find that benefits are realized for both objectives compared with the baseline policy that synchronizes all controller pairs at equal rate.

\end{enumerate}

The rest of the paper is organized as follows. Section \ref{section:consistency}  formulates and solves the synchronization problem for the consistency level maximization objective (Obj. 1). In Section \ref{section:performance}, we present our emulation results and our learning algorithm for maximizing the network application performance (Obj. 2). Section \ref{section:evaluation} presents the evaluation of our proposed algorithms, while Section \ref{section:related} reviews our contribution compared to related works. We conclude our work in Section \ref{section:conclusion}.

\section{Maximizing Consistency Level} \label{section:consistency}

In this section, we show how to optimize the first objective, i.e., the consistency level. We begin by describing the system model and problem formulation.

\subsection{Model and Problem Formulation}

We adopt a general model representing an eventually-consistent SDN system with a set $\mathcal{C}$ of $C$ controllers distributed in a network, as shown in Figure \ref{fig:example}. Each controller is responsible for managing a subset of the data plane nodes in the network, referred to as a domain. The controllers are aware of the current state information inside their domains (e.g., available links, flow table entries). This can be achieved by using a SouthBound protocol (e.g., OpenFlow) for signaling and statistic collection from the data plane nodes.

The domain states may change dynamically as nodes and links fail or recover and new data flows are generated. To model such dynamics, we denote by $\lambda_i$ the rate of state changes in the domain of controller $i$. Specifically, we assume that the state changes follow an \emph{independent Poisson process} at rate $\lambda_i$. This rate can be predicted by the network operator for a certain time period (e.g., a few hours). The above assumptions, made for model tractability, will be relaxed in next section.

We divide the time period into slots (e.g., a few tens of seconds each). To ensure a \emph{minimum level of consistency}, in the beginning of each slot every controller disseminates a synchronization message conveying the current state of its domain to every other controller. By the end of a slot, the state of a controller $i$ might change or remain the same. According to the Poisson process model, the probability that the state of controller $i$ remains the same is $e^{-\lambda_{i} s }$, where $s$ is the time slot length. Therefore, with the above probability the state of controller $i$ remains consistent with the view that any other controller $j$ has on it. We note that, for a given pair of controllers $(i,j)$, it might happen that the state of controller $i$ is consistent with the respective view of $j$ but not the other way around. In this case, we say that only the controller pair $(i,j)$ is consistent, but not the controller pair ($j,i$). Overall, the expected number of controller pairs that are consistent  (\emph{consistency level}) is given by:
\begin{equation}
\sum_{i \in \mathcal{C}} \sum_{j \in \mathcal{C}, j\neq i}  e^{-\lambda_{i} s }
\end{equation}

\begin{figure}[t]
	\begin{center}
		\includegraphics[scale=0.7]{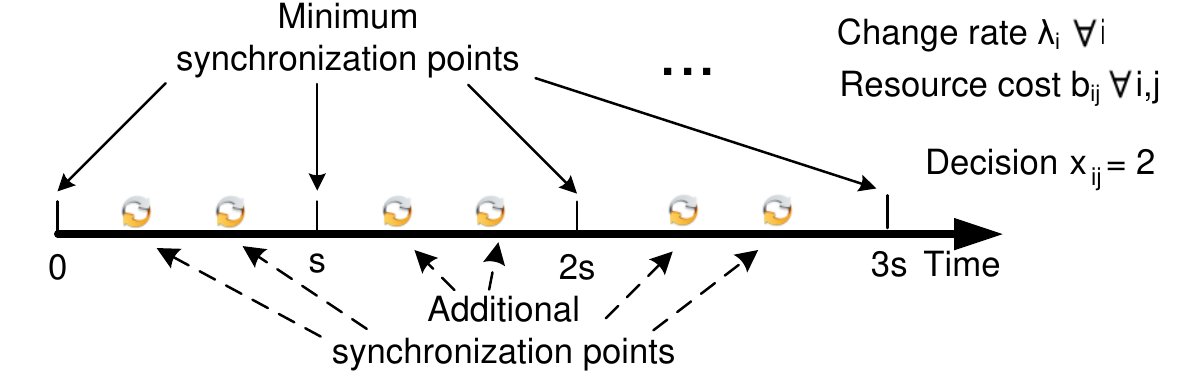}
		\caption{Overview of synchronization decisions.}
		\label{fig:obj1}
	\end{center}
\vspace{-2mm}
\end{figure}

To improve the consistency level, the controllers have the option to disseminate additional synchronization messages within each slot, as illustrated in Figure \ref{fig:obj1}. We denote by $x_{ij} \in \{0,1,\dots,R\}$ the number of such messages sent from controller $i$ to $j$, where $R$ represents the maximum possible synchronization rate. We note that the synchronization rates may be asymmetric in general, i.e., it may happen that $x_{ij} \neq x_{ji}$. A \emph{synchronization policy} can be expressed by the respective vector:
\begin{equation}
\bm{x} = (x_{ij} \in \{0,1,\dots,R\}~:~\forall i, j \in \mathcal{C}, j\neq i) \label{eq:x}
\end{equation}

The dissemination of synchronization messages is not without cost. It consumes network resources such as bandwidth and energy that can be significant especially in wireless resource-constrained environments with in-band control channels of limited capacity. The system operator has to ensure that certain resource constraint is met. Specifically, we require that:
\begin{equation}
 \sum_{i \in \mathcal{C}} \sum_{j \in \mathcal{C}, j\neq i} x_{ij} b_{ij} \leq B \label{eq:budget}
\end{equation}
where $b_{ij}$ is the resource cost of message dissemination between controllers $i$ and $j$,  which typically depends on the distance between the two controllers. $B$ is a positive constant representing the available network resources.

The disseminated synchronization messages will improve the consistency level of the system. Specifically, by spreading $x_{ij}$ messages uniformly over the slot interval $s$, we can effectively reduce the interval length by a factor of $x_{ij}+1$ (as illustrated in Figure \ref{fig:obj1}). Therefore, the probability that the controller pair $(i,j)$ is consistent increases from $e^{-  \lambda_{i} s}$  to $e^{- \frac{ \lambda_{i} s }{ x_{ij}+1 }}$.  The consistency level will be:
\begin{equation}
 \Omega (\bm{x}) =  \sum_{i \in \mathcal{C}} \sum_{j \in \mathcal{C}, j \neq i}  e^{- \frac{ \lambda_{i} s }{ x_{ij}+1 } }
\end{equation}

The objective of the system operator is to find the synchronization policy that maximizes the consistency level while satisfying the resource constraint:
\begin{eqnarray}
{\bf{Obj.~1:~}} \max _{ \bm{x}  }  &  \Omega(\bm{x}) \\
s.t. & \text{constraints:}~(\ref{eq:x}), (\ref{eq:budget})\nonumber
\end{eqnarray}


\subsection{Complexity and Solution}

We first prove the intractability of the problem.

\begin{theorem}
The SDN synchronization problem for Obj. 1 is NP-Hard. \label{theorem:1}
\end{theorem}
\begin{proof}
We prove the NP-Hardness of our problem by reduction from the \emph{Knapsack problem} (which is NP-Hard), defined as follows:
Given a knapsack of capacity $W$, and a set of $L$ items with nonnegative weights $w_1$ to $w_L$ and values $v_1$ to $v_L$, the objective is to place in the knapsack the subset of items such that the total value $V$ is maximized without the total weight exceeding $W$.

Every instance of the knapsack problem can be written as a special case of our problem  where: (i) we set $B=W$, (ii) we restrict $R=1$ and (iii) we create one controller pair ($i_l,j_l$) for each item $l$. Each such controller pair $(i_l,j_l)$ has cost $b_{i_lj_l}$ equal to the weight of the mapped item $w_l$. The $\lambda_{i_l}$ rate is set such that the difference $  e^{- \frac{ \lambda_{i_l} s }{ 2 } } - e^{-  \lambda_{i_l} s  } $ is equal to the value of the mapped item $v_l$. Any other pair of controllers ($i_l,j_{l'}$) where $l \neq l'$ is excluded by setting $b_{i_l,j_{l'}}=+\infty$.

Given a solution to our problem of consistency level $V + \sum_{i\in\mathcal{C}}\sum_{j\in\mathcal{C}, j \neq i} e^{- \lambda_{i} s }$, we can find a solution to the knapsack problem of total value $V$ by placing in the knapsack the items corresponding to the pairs of controllers that synchronized with each other.
\end{proof}

Next, we identify a connection of our problem to the following variant of the knapsack problem:
\begin{definition}
\emph{Multiple-Choice Knapsack (MCK)}: Given $K$ classes $E_1$, $E_2$,\dots,$E_K$ of items to pack in a knapsack of capacity $W$, where the $l^{th}$ item in class $E_k$ has weight $w_{kl}$ and value $v_{kl}$, choose \emph{at most one} item from each class such that the total value $V$ is maximized without the total weight exceeding $W$. \label{definition:1}
\end{definition}

\begin{figure*}[t]
	\centering
	\subfloat[]{
		\includegraphics[width =4.0cm, height=4.0cm]{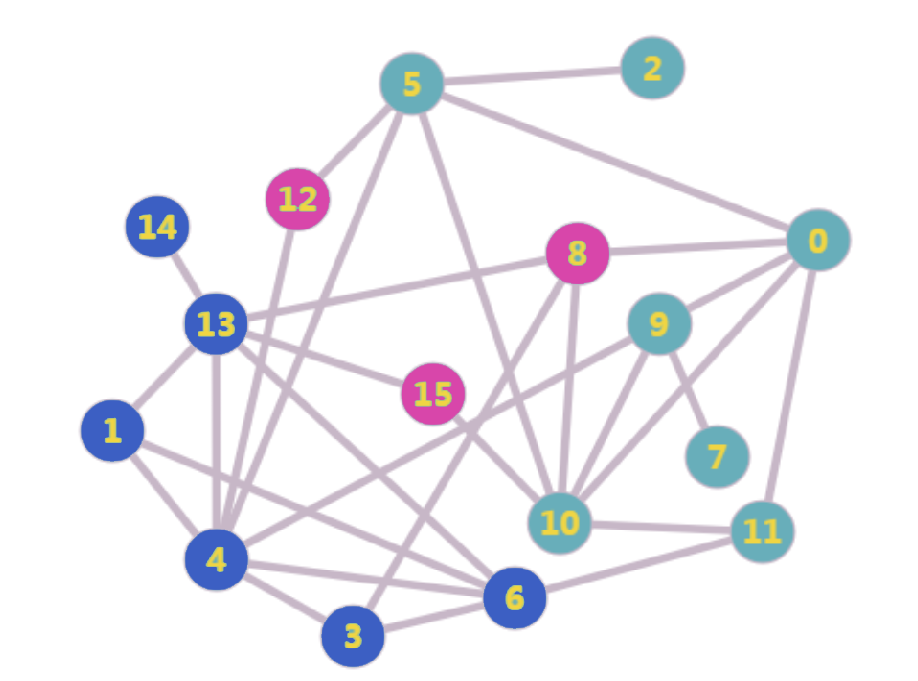} 	
		\label{fig:emulation1}
	}
	\subfloat[]{
		\includegraphics[width =4.75cm, height=4.0cm]{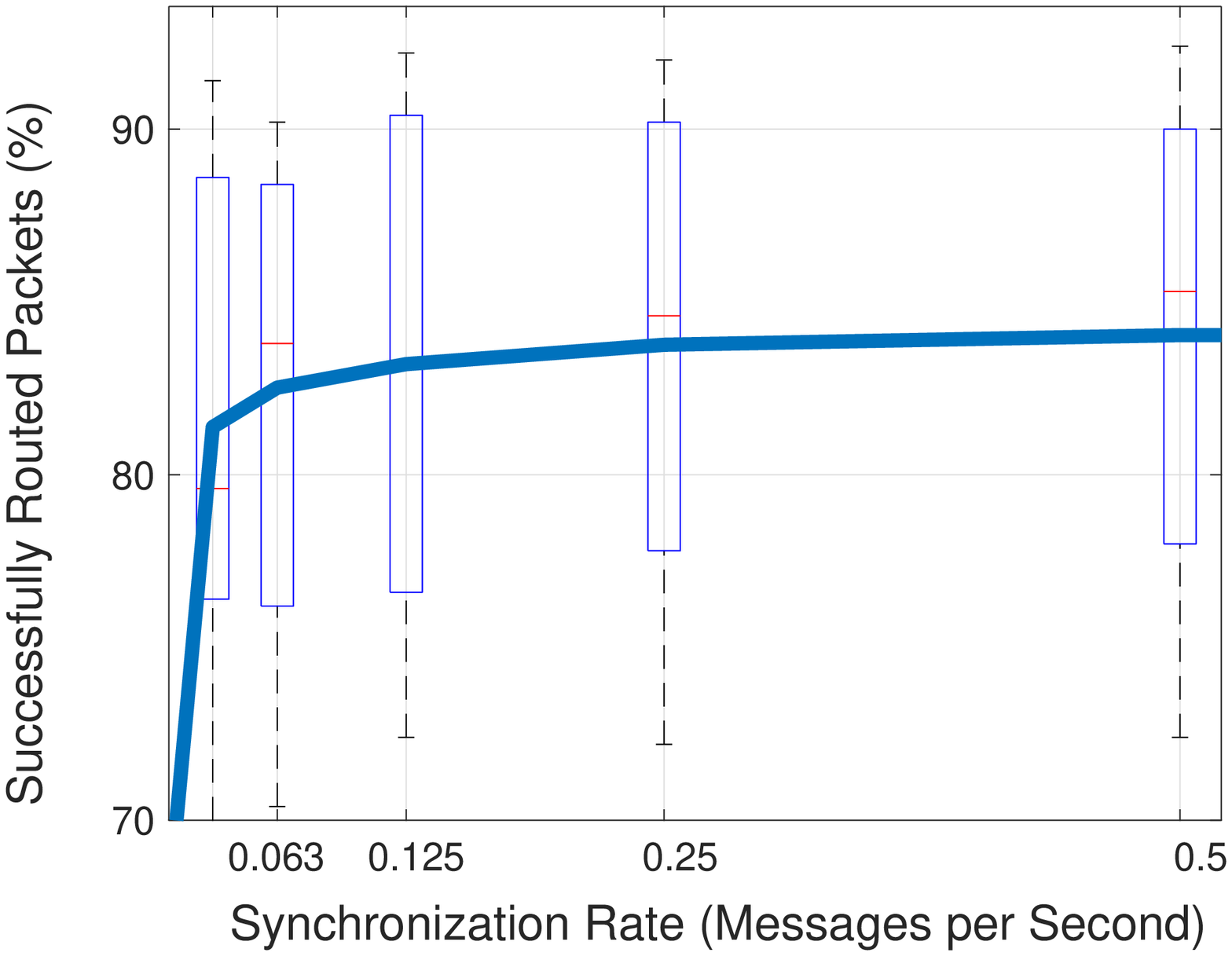} 			\label{fig:emulation2}
	}
	\subfloat[]{
		\includegraphics[width =4.0cm, height=4.0cm]{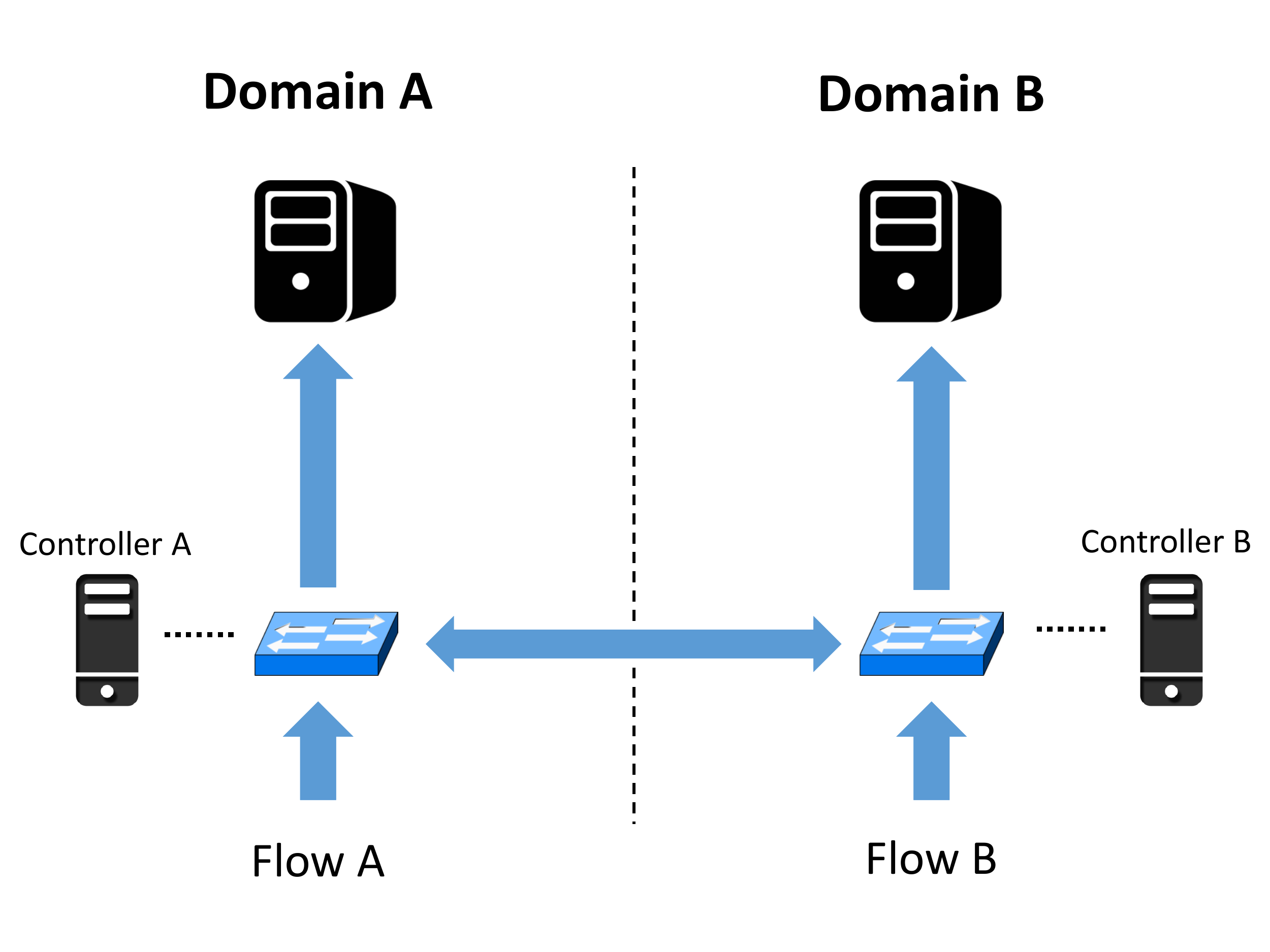}
		\label{fig:emulation3}
	}
	\subfloat[]{
		\includegraphics[width =4.75cm, height=4.0cm]{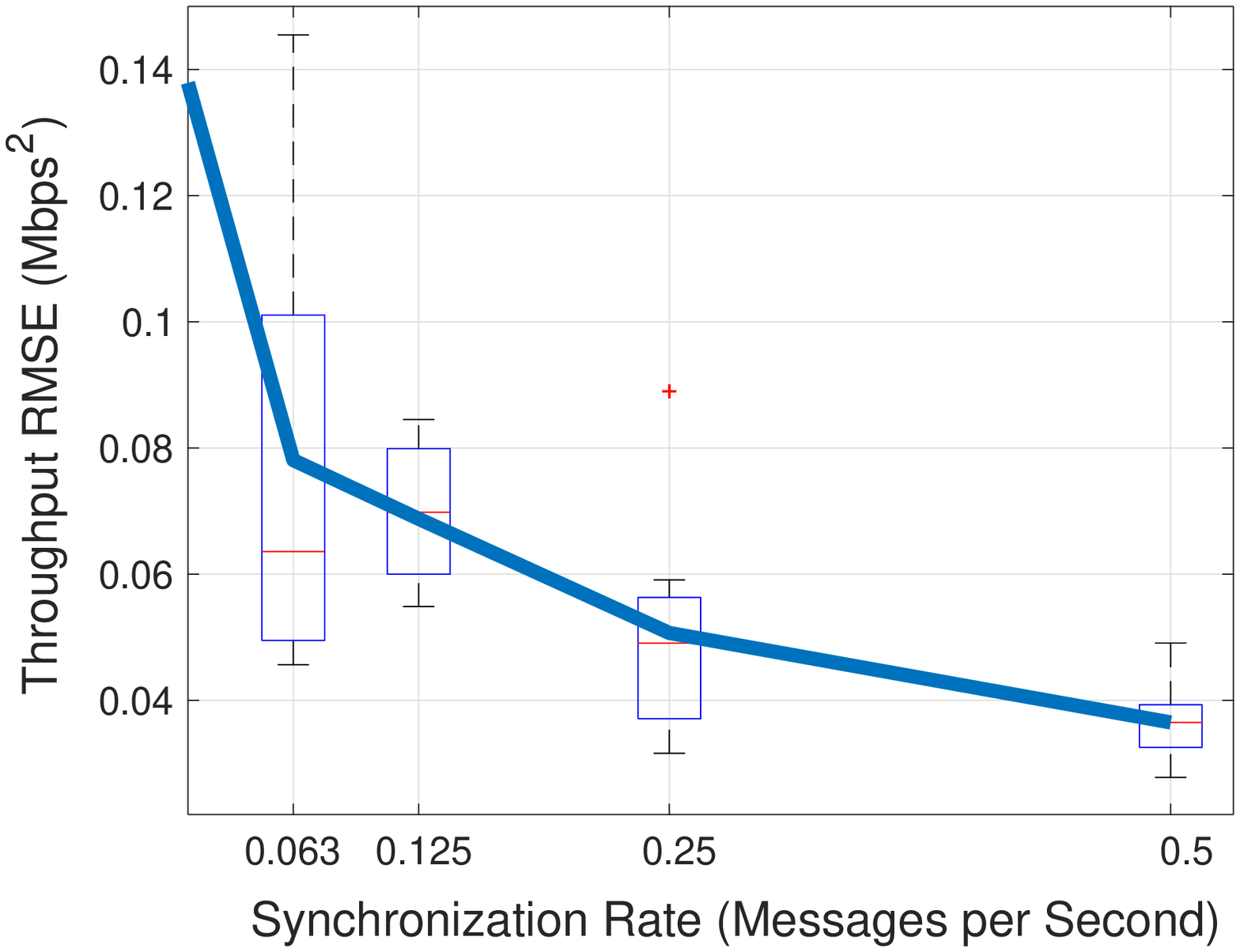} 			\label{fig:emulation4}
	}
	\caption{Emulation results. Topology and impact of synchronization rate on the performance (box plots and average values) of (a)(b) shortest path routing and (c)(d) load balancing applications.}
		\label{fig:emulation}
\end{figure*}

Specifically, the following lemma holds:
\begin{lemma}
The SDN synchronization problem for Obj. 1 is polynomial-time reducible to the problem MCK. \label{lemma:1}
\end{lemma}
\begin{proof}
Given an instance of the synchronization problem, we construct the equivalent instance of the problem MCK as follows:
We create a knapsack of size equal to $B$ and an item class $E_{k}$ for each pair of controllers $k=(i,j)$. Each class contains $R$ distinct items. The $l^{th}$ item in class $k$ has weight $w_{kl} =  b_{ij} l$ and value $v_{kl} = e^{- \frac{ \lambda_{i} s }{ l+1 } } - e^{-  \lambda_{i} s  }$.

Each solution of value $V$ to the MCK instance can be mapped to a solution to the synchronization problem instance of value $V+\sum_{i\in\mathcal{C}}\sum_{j\in\mathcal{C}, j \neq i} e^{- \lambda_{i} s }$ as follows:
If the $l^{th}$ item in class $E_k$ (where $k=(i,j)$) is packed in the knapsack, we synchronize controllers $i$ and $j$ at rate $x_{ij}=l$.
Clearly, the obtained solution spends no more resources than $B$. The value of the $l^{th}$ item $v_{kl}$ is equal to the increase in the consistency level due to the synchronization decision $x_{ij}=l$.
\end{proof}

Lemma \ref{lemma:1} is very important since it allows us to exploit a wide range of solution algorithms that have been proposed for problem MCK to solve our problem. In particular, although MCK is NP-hard, \emph{pseudopolynomial-time optimal} and \emph{fully-polynomial-time approximation} (FPTA) algorithms are known~\cite{mck}. By pseudopolynomial we mean that the running time is polynomial in the input (knapsack capacity and item weights), but exponential in the length of it (number of digits required to represent it). The FPTA algorithm ensures that the performance of the solution is no less than $(1-\epsilon)$ fraction of the optimal, while its running time is polynomial to $\frac{1}{\epsilon}$, $\epsilon \in (0,1)$. Therefore, the running time and performance of FPTA are adjustable, making it preferable for large problem instances. Hence, we obtain the following result:

\begin{theorem}
There exists a pseudopolynomial-time optimal algorithm and a FPTA algorithm to the SDN synchronization problem for Obj. 1.
\end{theorem}

\section{Maximizing Application Performance} \label{section:performance}

Our work in the previous section constitutes the first systematic approach to tackle the SDN synchronization problem. However, it has two limitations. First, it relies on the assumption that the state dynamics follow a specific distribution (independent Poisson with known rate $\lambda_i$). Second, while the consistency level (Obj. 1) is an important indicator of the performance of network application, it may not be always accurate. In fact, it is known that certain applications can tolerate some inconsistency among controllers provided that conflicts are avoided, while other applications have stricter requirements~\cite{levin}, \cite{scl}. The above motivate us to look for alternative synchronization methods that (i) \emph{are agnostic to the distribution of state dynamics} and (ii) \emph{optimize directly the performance of specific applications of interest rather than the consistency level}. In this section, we describe such a method by leveraging elements from the learning theory. Before that, we provide a brief  emulation study that will highlight the impact of synchronization rate on the performance of some popular network applications.


\subsection{Emulation Study} \label{subsection:emulation}

Below, we describe the emulation setup that will be later used to test the performance of two network applications, namely shortest path routing and load balancing.

\textbf{Emulation setup}.
We use \emph{Mininet}~\cite{mininet} to emulate virtual networks with several nodes and SDN controllers running on the same CPU machine. Among the set of commercial controllers that are available online we pick \emph{RYU}~\cite{ryu} which is open-source and allows us to develop our own protocols for the synchronization among controllers. Specifically, we implement a simple eventually-consistent protocol which periodically disseminates synchronization messages between each controller pair. Our code is parameterized to allow for any synchronization period. The disseminated messages convey the local views of controllers about the topology and installed flow tables. This information is made available to the controllers by the OpenFlow protocol.

\textbf{Emulation results}.
We first test the performance of a shortest path routing application. With this application, packets are routed to their destination following the path of minimum hop count, calculated by Dijkstra's algorithm. We generate the random network of 16 nodes and 3 controllers, depicted in Figure \ref{fig:emulation1}, where links fail or recover randomly and independently every one second with probability $0.05$, and nodes with the same color are managed by the same controller. We further generate data packets with random source-destination nodes. Unless the controllers synchronize at the time of packet generation, the packet is at risk of following a failed routing path.

The performance of routing application is determined by the number of packets that are successfully routed (without traversing any failed link) to their destinations. We emulate the performance for five different scenarios where all the controller pairs synchronize at the same rate equal to (i) 0.5, (ii) 0.25, (iii) 0.125, (iv) 0.063 and (v) 0.031 (messages per second). This translates to a single message disseminated every 2, 4, 8, 16 or 32 seconds. For each scenario, emulations are run for multiple times and the results are depicted in Figure \ref{fig:emulation2}. Despite a large extent of randomness, we observe that the average performance (calculated over $20$ minutes) increases with the synchronization rate and saturates eventually showing that a \emph{diminishing return rule} applies.

We perform additional emulations to test the performance of a load balancing application. We consider a similar setup with the work in \cite{levin}, depicted in Figure \ref{fig:emulation3}. That is, we generate a network with two controllers. Each controller manages two nodes, a switch and a server. The switches generate flows uniformly at random. The flows can be routed and queued to any of the two servers. Each controller is aware of the load of the server it manages. It also receives periodic synchronization messages about the load of the other server by the other controller. Each time a new flow is generated, the responsible controller routes it to the server with the currently observed lowest load. However, this may not be the least loaded server in reality, since the controllers are not synchronized at all times.

The emulation results are depicted in Figure \ref{fig:emulation4}. The metric we consider is the root-mean-square deviation (RMSE) of two servers' throughputs. The better the two server loads balance, the lower the value of this metric becomes. Therefore, this metric captures the performance of a load balancing application. For convenience, we claim it the cost function, and denote the performance metric the opposite value of cost function. Then, coinciding with the routing application, we observe that the performance improves with the synchronization rate but gradually saturates showing that a diminishing return rule applies.

\subsection{Learning Framework}

Subsequently, we study the objective of maximizing the performance of a network application (Obj. 2) such as the applications emulated in the previous subsection. While the objective function is expected to have a curve shape similar to those reported in Figure \ref{fig:emulation}, we cannot express in closed-form how exactly the synchronization rates will affect application performance. Therefore, the objective function is \emph{unknown}, rendering the problem fundamentally different (and more challenging) than Obj. 1.

To overcome the unknown objective challenge, we propose to leverage methods from the  \emph{learning theory}. Such methods typically \emph{train} the system by trying-out a sequence of solutions (synchronization rates) over some training period $\mathcal{T} = \{1,2,\dots,T\}$ of $T$ time slots, until they can infer a ``sufficiently good'' solution. To describe such training process, we generalize the vector of synchronization rate variables as:
\begin{equation}
\bm{x} = (x^t_{ij} \in \{0,1,\dots,R\}~:~\forall i, j \in \mathcal{C}, j\neq i, t\in \mathcal{T}) \label{eq:x2}
\end{equation}
where $x^t_{ij}$ indicates the synchronization rate between controllers $i$ and $j$ tried-out in time slot $t$. We further denote by the vector $\bm{x}^t$ all the variables in time slot $t$. We emphasize that the variable values will be typically different from slot to slot as different synchronization rates need to be explored in order to train the system.

Given the synchronization rate vector $\bm{x}^t$ tried-out in a slot $t$, the application performance will be $\Psi_t({\bm{x}^t})$. Here, $\Psi_t(.)$ is an unknown function that governs the application performance in slot $t$. While the overall function is unknown, the single value $\Psi_t({\bm{x}^t})$ can be \emph{observed} by the system operator \emph{after} the synchronization rate decision ${\bm{x}^t}$ is made, in the end of the slot. For a shortest path routing application, for example, this is possible by measuring the number of data packets that reached their destination in time. Such information can be estimated by the controllers using the TCP acknowledgement packets. The information can be then passed to the system operator (e.g., one of the controllers) which can simply aggregate and sum the respective values.

We emphasize that the function $\Psi_t(.)$ is time slot-dependent, meaning that the performance value might change with time even for the same synchronization rate decision. That is, we may try-out the same synchronization rate vector $\bm{x}^{t} = \bm{x}^{t'}$ in two slots $t$ and $t'$ but observe different performance values $\Psi_{t}(\bm{x}^{t}) \neq \Psi_{t'}(\bm{x}^{t'})$.
Such \emph{uncertainty of observations} is due to the stochastic nature of the network. Intuitively, the performance value will be large if the network happens to be stable in a slot but will be much worse in other slots during which many changes happen.

Despite the uncertainty of observations, the learning method should be able to infer by the end of the training period $\mathcal{T}$ a ``sufficiently good'' synchronization rate decision $\widehat{\bm{x}} = ( \widehat{x}_{ij} : i,j \in \mathcal{C}, j\neq i )$. This should, ideally, maximize the \emph{average} performance denoted by an (also unknown) function $\widehat{\Psi}(.) = \mathbb{E}[\Psi_t(.)]$. While the system operator does not know the average performance values, we assume that they do not change over a period of time (e.g., a few hours). Therefore, our second objective can be written as:

\begin{eqnarray}
{\bf{Obj.~2:~}} \max _{ \widehat{\bm{x}}  }  &  \widehat{\Psi}( \widehat{\bm{x}} ) \\
s.t. &  \sum_{i\in\mathcal{C}} \sum_{j\in\mathcal{C}, j \neq i} \widehat{x}_{ij} b_{ij} \leq B \label{eq:budget2}
\end{eqnarray}
where inequality (\ref{eq:budget2}) ensures that the inferred synchronization rate decision will satisfy the resource constraint.

We need to emphasize that the average performance $\widehat{\Psi}( \widehat{\bm{x}} )$ can be in fact the aggregate of many (rather than only one) applications. Either way, the performance is not the only criterion that determines the efficiency of a learning method. Another important criterion in this context is the \emph{running (or training) time} $T$, i.e., how many time slots are required for training in order to infer the synchronization rate decision $\widehat{\bm{x}}$. In the next subsection, we will propose a learning method that has adjustable average performance and running time.

\subsection{Learning Algorithm}

To handle the uncertainty of an observed performance value $\Psi_t(\bm{x}^t)$, a learning method would typically try-out the \emph{same} synchronization decision $\bm{x}^t$ \emph{multiple times}, in different time slots. Then, the empirical mean of the observations will be used to estimate the average performance value $\widehat{\Psi}(\bm{x}^t)$.
By repeating the above training process for every possible synchronization decision,  an estimate of the entire objective function $\widehat{\Psi}(.)$ can be obtained. However, there exists an \emph{exponential number of possible decisions}; $(R+1)^{C(C-1)}$ decisions in total. Therefore, this approach would require an exponential number of time slots for training, which is clearly not practical.

To overcome the high dimensionality of the synchronization decision space, we could leverage learning methods proposed recently that do not require the estimation of the objective function at every possible decision. For instance, the \emph{ExpGreedy algorithm} proposed in \cite{sub-noisy} can infer a close-to-optimal decision in \emph{polynomial-time} provided that the objective function follows a diminishing return rule, as the one observed in the emulation results in Figure \ref{fig:emulation}. Still, however, the running time of this algorithm may be too large for our problem, as we will show numerically in the next section, hindering its application in practical scenarios.

Based on the above, we propose an alternative more-practical learning algorithm for which we can flexibly adjust the running time by setting appropriate values to its input parameters. We refer to this algorithm as \emph{Stochastic Greedy} and summarize it in Algorithm 1. To ease presentation, we have assumed that the resource costs are equal and normalized to one for all the controller pairs, i.e., $b_{ij}=1$ $\forall i,j$. However, the algorithm and analysis can be easily extended for heterogeneous resource costs.

In a nutshell, the Stochastic Greedy algorithm starts with the all-zero synchronization decision and then gradually constructs the decision to be returned by iteratively increasing by $1$ the synchronization rate of a single controller pair. This procedure will end when the $B$ resource constraint is reached, i.e., after $B$ iterations. Each iteration requires multiple time slots for training so as to be confident that the controller pair selected to increase its rate by $1$ will improve the average performance more than other controller pairs. The length of the training period can be adjusted by two input parameters $\sigma$ and $\tau$. The value of $\sigma$ is between $1$ and $C(C-1)$, while $\tau$ can take any positive integer value.

Formally, the algorithm maintains a synchronization rate decision $\widehat{\bm{x}}$, initially set to the zero vector $\bm{0}$ (line \ref{alg2:1}). It spends the first $\tau$ time slots trying out the zero synchronization decision and uses the $\tau$ observations to estimate  $\widehat{\Psi}(\bm{0})$ (lines \ref{alg2:2}-\ref{alg2:3}). In the next $B$ iterations (lines \ref{alg2:loop_start}-\ref{alg2:11}), the algorithm will iteratively select a controller pair and increase the respective synchronization rate by $1$, updating $\widehat{\bm{x}}$. At each iteration $k=1,2,\dots,B$, the algorithm will initially pick $\sigma$ random pairs of controllers as candidates (line \ref{alg2:candidates}). For each such pair $p=1,2,\dots,\sigma$, the synchronization decision $\widehat{\bm{x}}'$ will be set accordingly (line \ref{alg2:x'}) and $\tau$ time slots will be spent to estimate $\widehat{\Psi}(\widehat{\bm{x}}')$ (lines \ref{alg2:8}-\ref{alg2:9}). The marginal performance gain of switching from decision $\widehat{\bm{x}}$ to $\widehat{\bm{x}}'$, denoted by $D(\widehat{\bm{x}}, \widehat{\bm{x}} ')$,  will be estimated (line \ref{alg2:loop_end}). Among the $\sigma$ candidate controller pairs, the algorithm will include in the current decision $\widehat{\bm{x}}$ the pair with the maximum estimated marginal performance gain (line \ref{alg2:11}).


The algorithm spends $\tau$ time slots to estimate $\widehat{\Psi}(\widehat{\bm{x}})$ for $\widehat{\bm{x}}=\bm{0}$, and $\sigma \tau$ more slots for each iteration. Therefore, the total running (or training) time is $T= \tau + \sigma \tau B$ time slots. The following theorem describes the average performance of the algorithm. Since the algorithm makes random decisions, the average performance bound holds in expectation.

\begin{theorem}
Algorithm 1 achieves average performance $\widehat{\Psi}(\widehat{\bm{x}})$ that is in expectation a factor $1 - e^{-(1-\epsilon) \mu  }$ from the optimal where $\epsilon = e^{-\sigma \frac{B}{C(C-1)R}} $ and $ \mu $ is the expected fraction of the observed marginal gain in a slot over the actual marginal gain. \label{theorem:3}
\end{theorem}

To facilitate reading, we defer the proof of the theorem to the appendix. We emphasize that the average performance bound depends on the value of $\mu$. This value captures the uncertainty of the observed performance values since the changes in network state  may be unevenly distributed across the time slots. If $\mu=1$, it means that the performance value does not depend on the time slot of observation and hence the estimated maximum performance will be the actual one. However, as the $\mu$ value goes to $0$ the observations become more uncertain.

\begin{algorithm}[t]
\BlankLine
\nl Initialize $\widehat{\bm{x}}=\bm{0}$; \label{alg2:1} \\
\nl Try out $\bm{x}^{t} = \widehat{\bm{x}}$ and observe $\Psi_t(\bm{x}^t)$ $\forall t \in \{1,\dots,\tau\}$; \label{alg2:2}\\
\nl Estimate $\widehat{\Psi}(\widehat{\bm{x}}) = \frac{1}{\tau} \sum_{t=1}^{\tau} \Psi_t(\bm{x}^t)$; \label{alg2:3}\\
\nl \For{each iteration $k$ from $1$ to $B$}{ \label{alg2:loop_start}
\nl Pick $\sigma$ random controller pairs $p$ for which $\widehat{x}_{p}<R$; \label{alg2:candidates}\\
\nl \For{each picked pair $p$ from $1$ to $\sigma$}{
\nl Set ${\widehat{\bm{x}}'} = {\widehat{\bm{x}}}$ where $\widehat{x}'_p = \widehat{x}_p+1$;  \label{alg2:x'}\\
\nl Try out $\bm{x}^{t} = \widehat{\bm{x}}'$ and observe $\Psi_t(\bm{x}^t)$ $\forall t \in \{ (k-1)\sigma\tau + p\tau+1, \dots , (k-1)\sigma\tau + p\tau +\tau \}$;\label{alg2:8}\\
\nl Estimate  $\widehat{\Psi}(\widehat{\bm{x}}') = \frac{1}{\tau} \sum_{t= (k-1)\sigma\tau + p\tau +1 }^{(k-1)\sigma\tau + p\tau +\tau } \Psi_t(\bm{x}^t)$;\label{alg2:9}  \\
\nl Set $D(\widehat{\bm{x}}, \widehat{\bm{x}}' ) = \widehat{\Psi}(\widehat{\bm{x}}') - \widehat{\Psi}(\widehat{\bm{x}})$; \label{alg2:loop_end}\\
}
\nl Update $ \widehat{\bm{x}}  = \text{argmax}_{  \widehat{\bm{x}}'  }  D(\widehat{\bm{x}}, \widehat{\bm{x}}' ) $; \label{alg2:11} \\
}
\nl {\bf{Output}}: $\widehat{\bm{x}}$;
\caption{Stochastic Greedy with $(\sigma,\tau)$ input} \label{alg:sg}
\end{algorithm}

\begin{figure*}[t]
	\centering
	\subfloat[]{
	\includegraphics[width =5.5cm, height=4.0cm]{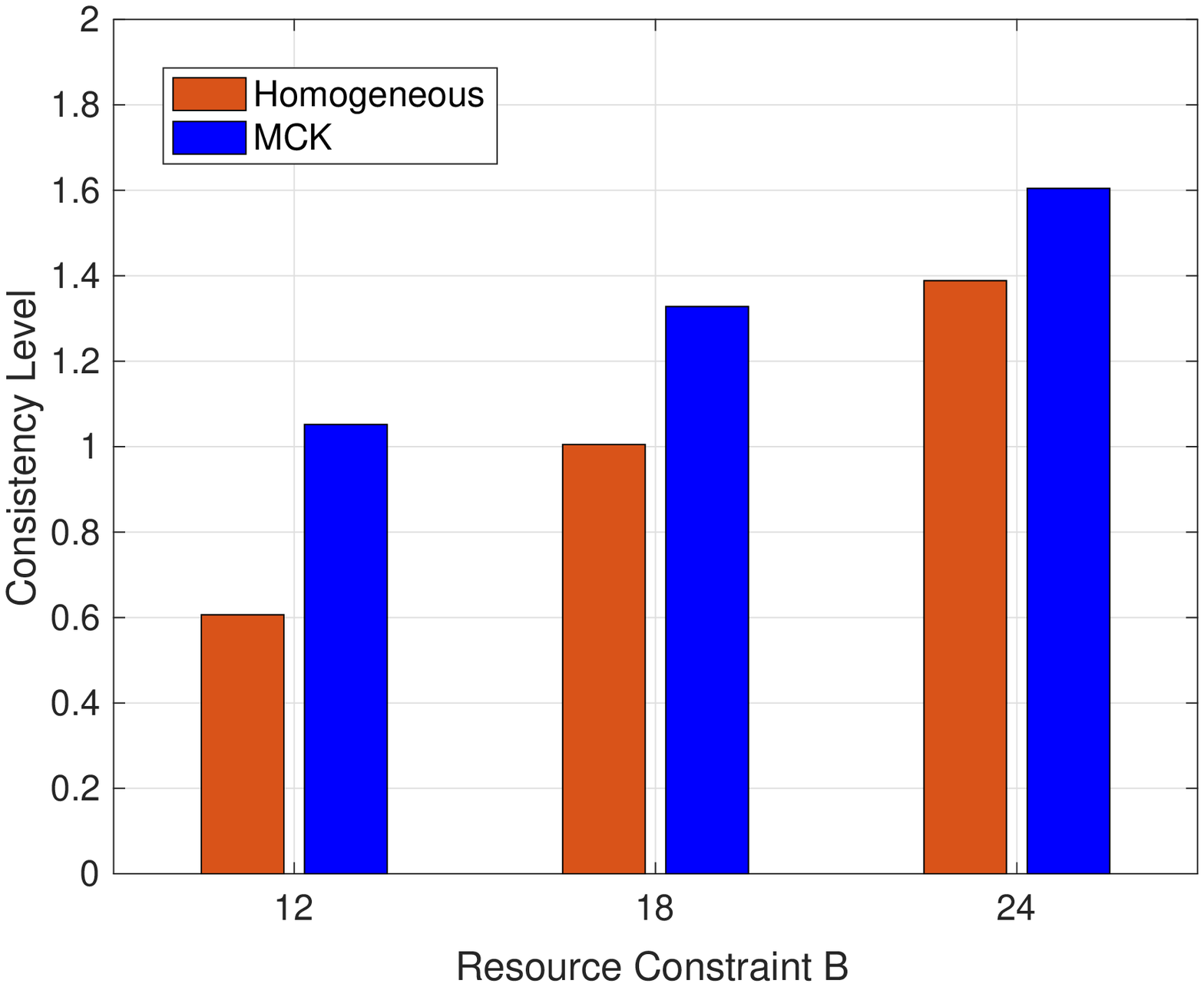} 		
	\label{fig:simulation1}
	}
	\subfloat[]{
	\includegraphics[width =5.5cm, height=4.0cm]{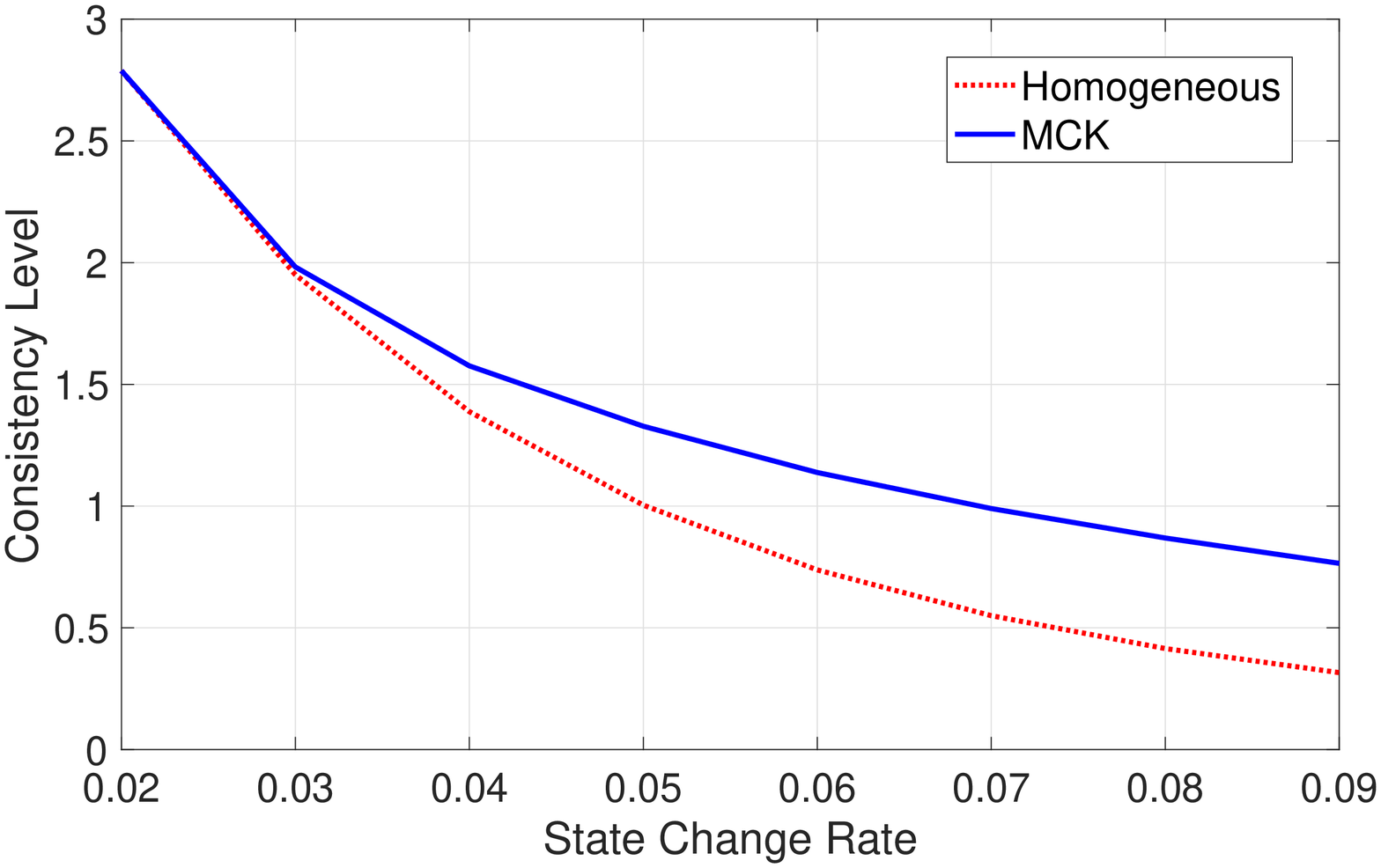} 		
	\label{fig:simulation2}
	}
    \subfloat[]{
	\includegraphics[width =5.5cm, height=4.0cm]{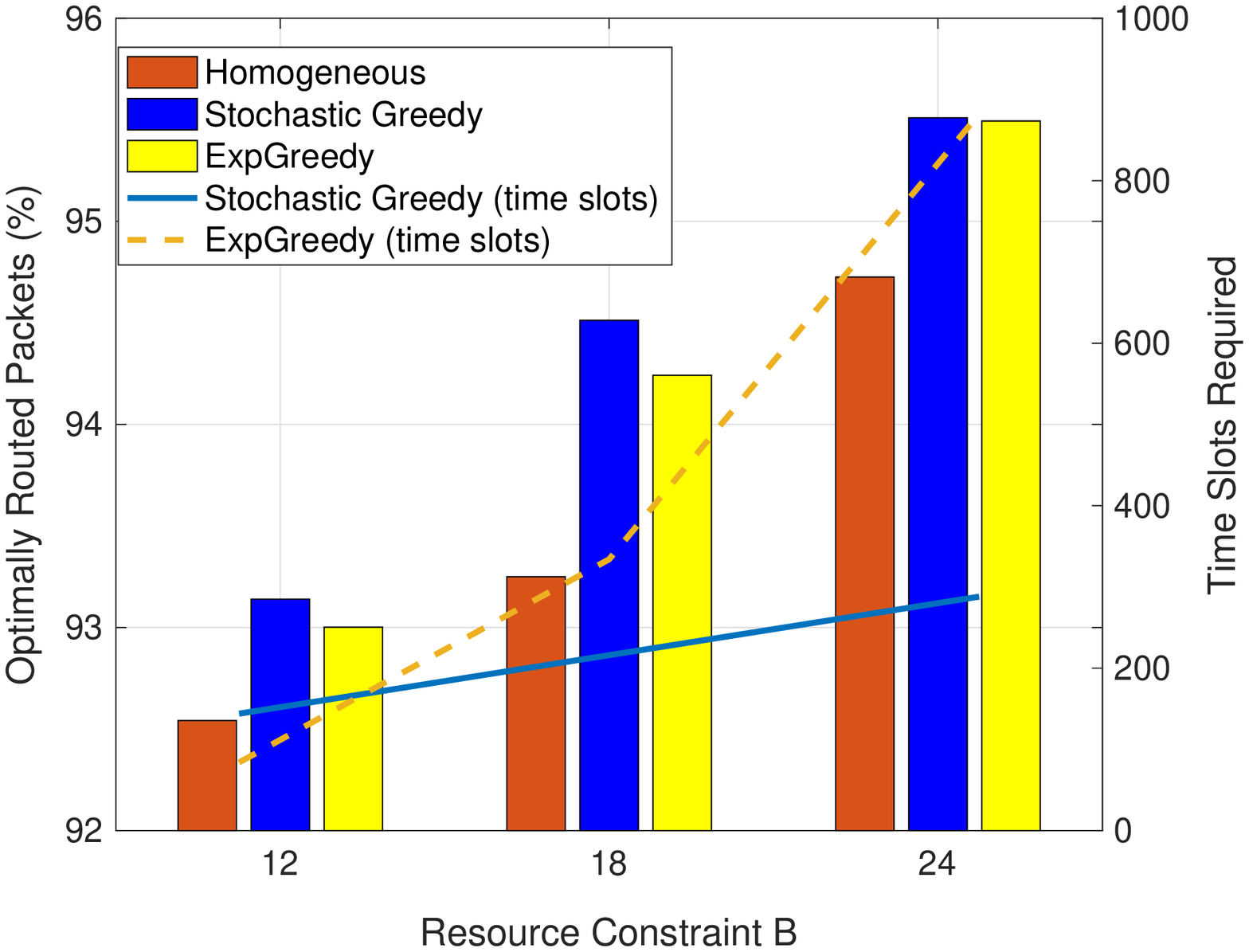} 		
	\label{fig:simulation3}
	}\\
	\subfloat[]{
	\includegraphics[width =5.5cm, height=4.0cm]{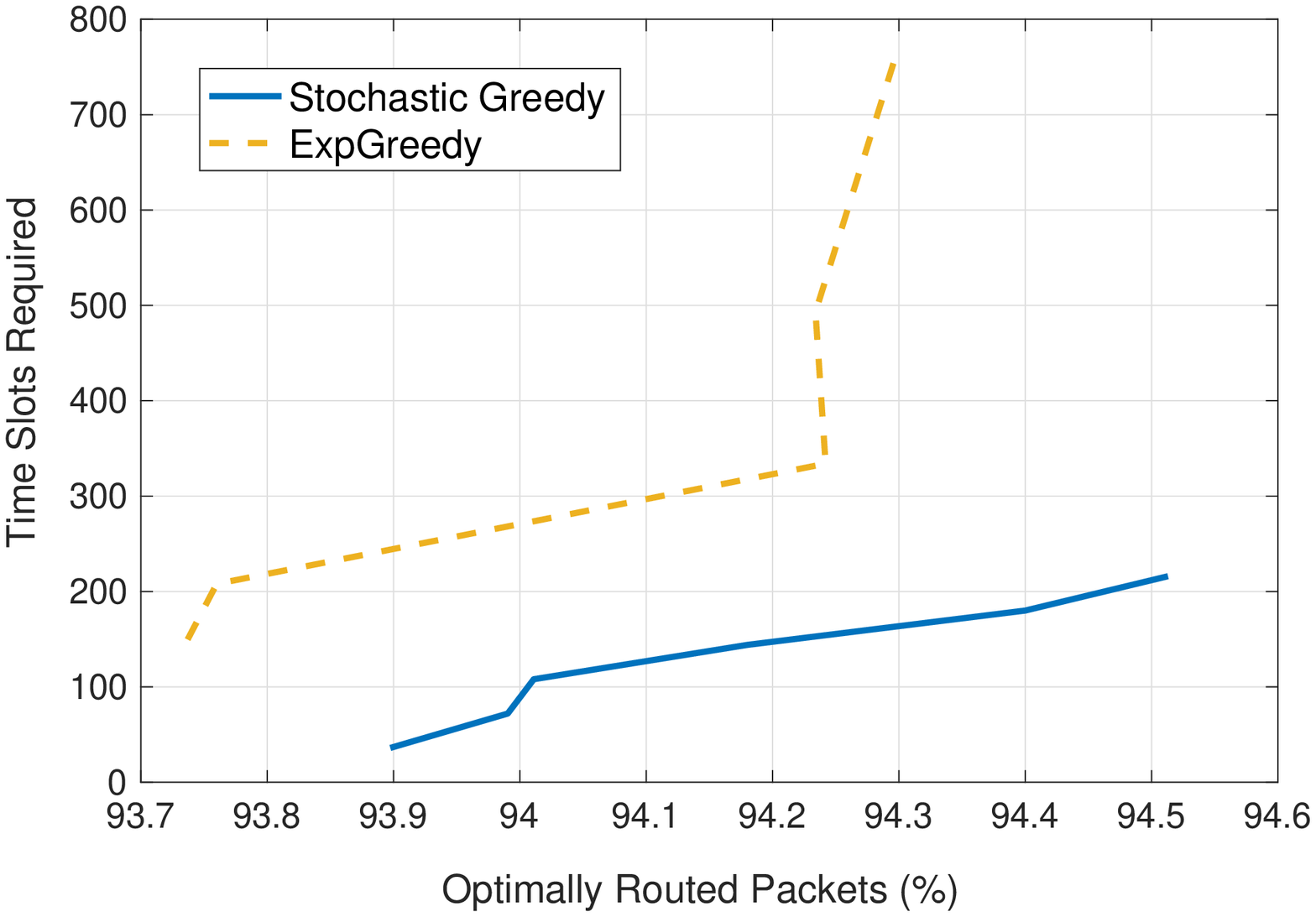} 	
	\label{fig:simulation4}
	}
	\subfloat[]{
	\includegraphics[width =5.5cm, height=4.0cm]{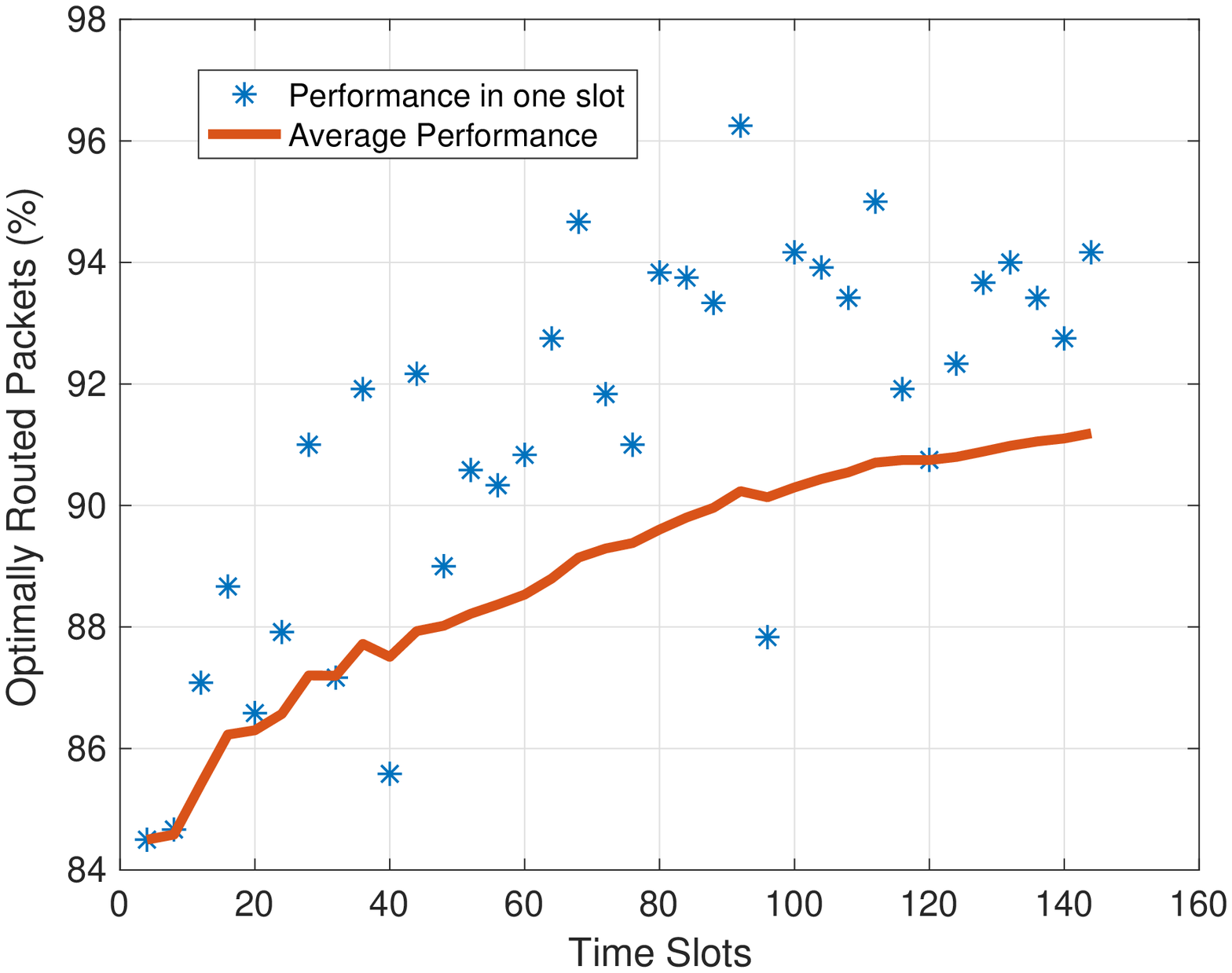} 		
	\label{fig:simulation5}
	}
	\subfloat[]{
		\includegraphics[width =5.5cm, height=4.0cm]{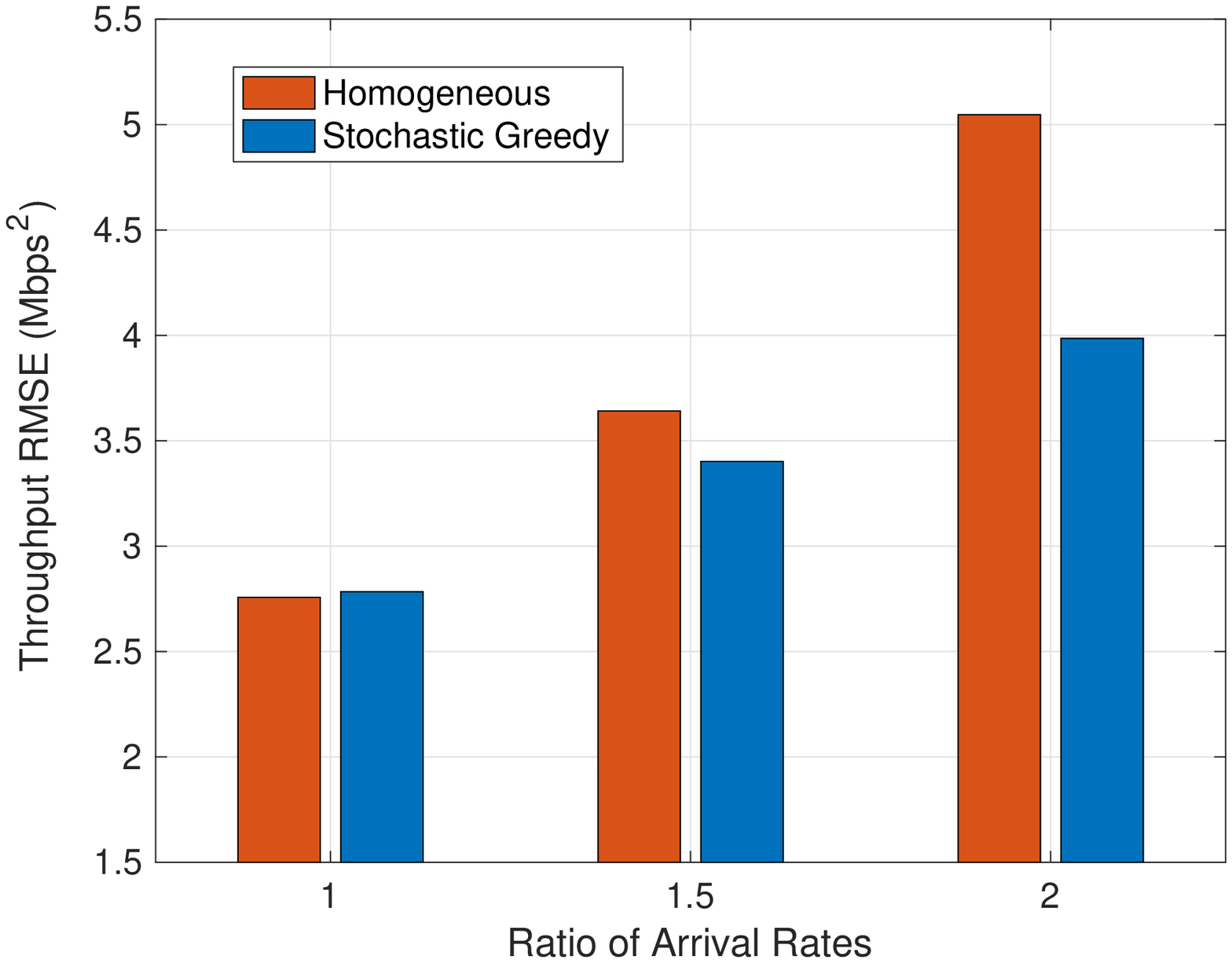} 		
		\label{fig:simulation6}
	}
	\caption{Consistency level for different (a) resource budgets and (b) rates of network state changes. (c) Performance and training time for different resource budgets, (d) tradeoff between performance and training time and (e) learning process under the shortest path routing application. (f) RMSE cost for different ratios of flow arrival rates under the load-balancing application.}
	\label{fig:simulation}
\vspace{-2mm}
\end{figure*}

Another issue is that the performance bound in Theorem \ref{theorem:3} holds in expectation, which means that it may be violated in practice. Therefore, it is important to bound the extent to which this happen, as we show in the following theorem.

\begin{theorem}
Algorithm 1 achieves average performance $\widehat{\Psi}(\widehat{\bm{x}})$ that is a factor $1 - e^{ -(1-\epsilon)(1-\gamma) \mu }$ from the optimal with probability $1-e^{-\frac{\gamma B \tau}{2}}$ for any $\gamma\in(0,1)$. \label{theorem:4}
\end{theorem}

The average performance bounds of our algorithm can be better understood through an example. In particular, consider the system with $C=5$ controllers, $B=10$ available resources and $s=30$ seconds per time slot. By picking $\sigma=5$ out of the $20$ possible controller pairs and $\tau=3$ time slots per try-out, the total running (training) time of the algorithm will be about one hour. Moreover, if the observed marginal performance gains are $50\%$  or more of the actual ones ($\mu=0.5$) and $R=1$, the average performance achieved by the algorithm will be in expectation at least $37\%$ of the optimal.
Picking a larger $\sigma$ value will increase the average performance (cf. Theorem \ref{theorem:3}). Picking a larger $\tau$ value will increase the probability that the performance bound is not violated (cf. Theorem \ref{theorem:4}).


\section{Evaluation Results} \label{section:evaluation}

In this section, we carry out evaluations to show the benefits of the proposed algorithms. Overall, we find that benefits are realized for both objectives compared with the baseline algorithm that synchronizes all the controller pairs at equal rate (referred to as Homogeneous). Especially for Obj. 2, our Stochastic Greedy algorithm achieves better performance-training time tradeoff than a state-of-the-art learning algorithm (ExpGreedy in \cite{sub-noisy}).

\textbf{Evaluation setup}.
We choose the same network topologies and applications as in our emulations in Section \ref{subsection:emulation} (16-node shortest path routing and 2-server load balancing). For Obj. 1, we compare the optimal algorithm according to our model (MCK) with the Homogeneous algorithm. For Obj. 2, we compare our Stochastic-Greedy algorithm with both the Homogeneous and ExpGreedy algorithms. To eliminate randomness, we run each algorithm 10 times and take the average value.


\textbf{Evaluation of Obj. 1}. We start with Obj. 1 and focus on the 16-node network with $C=3$ controllers (domains). We set the rate of changes of the $i^{th}$ domain as $\lambda_{i} = n_i \lambda$, where $n_i$ is the size of domain $i$, i.e., the number of data plane nodes, and $\lambda=0.05$. In Figure~\ref{fig:simulation1}, we calculate the consistency level of both MCK and Homogeneous algorithms for different resource budgets $B$. While the consistency level increases with $B$ for both algorithms, Homogeneous cannot reach the same level of consistency as MCK. Furthermore, in Figure~\ref{fig:simulation2} we investigate the impact of state change rate $\lambda$. In accordance with intuition, the more frequently changes happen, the more enhancement of consistency we can achieve by optimizing Obj. 1.

\textbf{Evaluation of Obj. 2: Shortest path routing application}. Next, we study Obj. 2, i.e., the performance of network applications. We first consider the shortest path routing application in the 16-node network. A performance metric of interest for this application is the percentage of packets that are optimally routed to their destinations, i.e., following paths of the same number of hops as the optimal path. Figure \ref{fig:simulation3} depicts the performance for different resource budgets $B$.  We notice that \emph{the proposed Stochastic-Greedy algorithm routes optimally more packets than Homogeneous and ExpGreedy algorithms}. The training time required by our algorithm increases linearly with $B$. On the other hand, the time of ExpGreedy increases more dramatically, which shows that our algorithm is more scalable. Specifically, \emph{our algorithm requires around 200 time slots (about an hour and a half) for training while ExpGreedy may consume more than 800 time slots (6-7 hours)}, which may be prohibitively large in practice.

The training time can be reduced for both algorithms by adjusting the input parameter values they take ($\sigma$ and $\tau$ for our algorithm). However, this will be at the cost of reduced performance (as we described in Theorems \ref{theorem:3} and \ref{theorem:4}). Figure~\ref{fig:simulation4} depicts the detailed tradeoff between performance and training time. It shows that we can flexibly trade the performance and training time of our algorithm (from $93.9\%$ to $94.5\%$ optimally routed packets and from $50$ to $210$ slots). For the same performance, ExpGreedy typically takes more than twice the time compared to our algorithm.
Figure~\ref{fig:simulation5} illustrates the learning process when Stochastic Greedy is run for $B=18$, $\sigma=2$ and $\tau=4$. Although in each time slot the algorithm observes a performance value with large randomness, it is able to allocate resources to proper pairs and increase the average performance over time.


\textbf{Evaluation of Obj. 2: Load balancing application}. Finally, we examine the load balancing application. Similar to the emulations in Section \ref{subsection:emulation}, we randomly generate flows at two switches. We define one time slot as $60$ seconds. Under the same $B$ value, we compare the Stochastic Greedy and Homogeneous algorithms for various flow arrival rates. When the arrival rates at the two switches are equal, the Homogeneous algorithm should be optimal because of the symmetry. In this case, as Figure~\ref{fig:simulation6} shows, our algorithm gets almost the same RMSE cost. Next, we set different arrival rates at the two switches. As a result, when the ratio of arrival rates gets larger, our algorithm leads to lower cost than the Homogeneous algorithm. For example, our algorithm can decrease the RMSE by around $20\%$ when the ratio of flow arrival rates at the two switches is equal to $2$.

\section{Related work} \label{section:related}

Distributed SDN controller deployments require a coordination protocol among controllers, which could easily generate significant amount of control traffic, e.g., see the measurement studies in~\cite{onos-traffic} and~\cite{infocom-18}. However, the control traffic is often neglected in literature with most of the existing works focusing on the routing and balancing of the data traffic, e.g., see~\cite{pavlou-tnsm} and the survey in \cite{sdn-survey}.

Some recent works proposed to reduce the overheads of control traffic by strategically placing the controllers in the network \cite{cp-mdcp} or by finding the appropriate forwarding paths for load balancing on control traffic~\cite{sdn-planning}. Nevertheless, the above approaches should be considered as complementary to our work, rather than competitive. On the one hand, the controller placement decisions are taken in a different (much slower) timescale than the synchronization. On the other hand, the control traffic forwarding can be combined with the synchronization rate decisions we make, since the former directly impacts the resource cost values $b_{ij}$ used as input to our problem.

Eventual consistency, where the controllers coordinate periodically rather than on demand basis, is another way to  reduce control overheads. Levin et al. \cite{levin} showed that certain network applications, like load-balancers, can work around eventual consistency and still deliver acceptable performance. This would require some additional effort to be made to ensure that conflicts such as forwarding loops, black holes and reachability violation are avoided~\cite{guo}.

Few recent works suggested the dynamic adaptation of synchronization period (or rate) among controllers in an eventually-consistent system so as to improve the performance of network applications while maintaining a scalable system~\cite{adaptive-consistency},~\cite{adaptive-consistency-credit}. While interesting and relevant, the above works did not provide any mathematical formulation or optimization framework. To the best of our knowledge our work is the first to systematically study the synchronization problem and propose optimization and learning methods.

\section{Conclusion} \label{section:conclusion}

In this paper, we studied the problem of finding the optimal synchronization rates among controllers in a distributed eventually-consistent SDN system. We considered two different objectives, namely, (i) the maximization of the number of controller pairs that are consistent, and (ii) the maximization of the performance of applications which may be affected by the synchronization decisions, as highlighted by emulations on a commercial SDN controller. For these objectives, we characterized the complexity of the problem and proposed algorithms to achieve the optimal synchronization rates. Evaluation results demonstrated significant performance benefits over the baseline policy that synchronizes all controller pairs at equal rate. Overall, the synchronization problem deserves more research attention, analogous to other problems in SDN framework. Our work in this paper can be seen as a kick-off for systematically studying optimization and learning methods to tackle this important problem.


\appendix[Proof of Theorem 3]

To facilitate the presentation of the proof, we describe an alternative representation of the synchronization rate decisions using the following set of elements (ground set):
\begin{equation}
\mathcal{G} = \{g^r_{ij} ~:~ \forall i,j\in C, j \neq i, r\in \{1,\dots,R\} \}
\end{equation}
Here, each of the elements $\{g^1_{ij}, g^2_{ij},\dots,g^{R}_{ij}\}$ indicates a separate message disseminated between controllers $i$ and $j$. Each subset of elements $\widehat{\mathcal{X}} \subseteq \mathcal{G}$ indicates a synchronization policy $\widehat{\bm{x}}$ where the synchronization rate $\widehat{x}_{ij}$ is equal to the number of the aforementioned elements included in $\widehat{\mathcal{X}}$.

We denote by the subsets $\mathcal{A} \subseteq \mathcal{G} $ and $\mathcal{O} \subseteq \mathcal{G} $ the solution returned by the Stochastic Greedy approximation algorithm and the optimal, respectively. We also denote by the subset $\mathcal{A}_k = \{\alpha_1,\dots,\alpha_k\} \subseteq \mathcal{A}$ the solution returned by the Stochastic Greedy algorithm after the first $0 \leq k \leq B$ iterations. Then, similar to the proof in \cite{sub-lazier}, we compute the probability that the set $\mathcal{S}$ of $\sigma$ elements that is randomly picked by Stochastic Greedy at iteration $k+1$ does not overlap with the optimal set $\mathcal{O}$ besides of the elements already in $\mathcal{A}_k$:
\begin{align}
\Pr[\mathcal{S} \cap (\mathcal{O}\setminus \mathcal{A}_k ) = \emptyset ]
& = \Big( 1-  \frac{|\mathcal{O}\setminus \mathcal{A}_k |}{|\mathcal{G}\setminus \mathcal{A}_k |} \Big) ^\sigma \nonumber \\
& \leq e^{-\sigma \frac{|\mathcal{O}\setminus \mathcal{A}_k |}{|\mathcal{G}\setminus \mathcal{A}_k |} } 
 \leq e^{-\sigma \frac{|\mathcal{O}\setminus \mathcal{A}_k |}{C(C-1)R} }
\end{align}
where the first inequality is because $(1-x)^a \leq e^{-ax}$ for any $x\in(0,1)$. The second inequality is because $|\mathcal{G}|=C(C-1)R$. Then, we have:
\begin{align}
\Pr[& \mathcal{S} \cap (\mathcal{O}\setminus \mathcal{A}_k ) \neq \emptyset ]
 \geq 1 - e^{-\sigma \frac{|\mathcal{O}\setminus \mathcal{A}_k |}{C(C-1)R} } \nonumber \\
&\geq ( 1 - e^{-\sigma \frac{B}{C(C-1)R} } ) \frac{|\mathcal{O}\setminus \mathcal{A}_k |}{B} 
= ( 1 - \epsilon ) \frac{|\mathcal{O}\setminus \mathcal{A}_k |}{B} \label{eq:probS1}
\end{align}
where the second inequality is because the function $1- e^{ -\sigma \frac{x}{C(C-1)R} }$ is concave with respect to $x \in [0,B]$. The last equality is because of the definition of $\epsilon$.

At iteration $k+1$, Stochastic Greedy adds the element $\alpha_{k+1}$ to the solution $\mathcal{A}_k$ which is estimated to maximize the marginal gain $\widehat{\Psi}(\mathcal{A}_{k+1})$ - $\widehat{\Psi}(\mathcal{A}_{k})$. However, the element with the real maximum marginal gain may be different, namely $\alpha'_{k+1} \neq \alpha_{k+1}$. Given that $\alpha_{k+1}$ is picked after $\tau$ try-outs, the  following equation holds:
\begin{align}
\widehat{\Psi}(\mathcal{A}_k & \cup\{\alpha_{k+1}\}) - \widehat{\Psi}(\mathcal{A}_k) =  \nonumber \\
&\Big( \sum_{t=1}^\tau  \frac{\mu^t_{k+1}}{\tau} \Big) \Big( \widehat{\Psi}(\mathcal{A}_k \cup\{\alpha'_{k+1}\}) - \widehat{\Psi}(\mathcal{A}_k) \Big)
\label{eq:mu1}
\end{align}
where $\mu^t_{k+1}$ is the ratio of marginal gains according to try-out $t=1,2,\dots,\tau$. Each $\mu^t_{k+1}$ value is taken from a distribution with mean value $\mu$.

By definition,  $\widehat{\Psi}(\mathcal{A}_k  \cup\{\alpha_{k+1}'\}) - \widehat{\Psi}(\mathcal{A}_k)$ is at least as much as the marginal value of an element randomly chosen from the set $\mathcal{S} \cap (\mathcal{O}\setminus\mathcal{A}_k)$ (if non-empty). This is actually an element randomly chosen from the entire set $\mathcal{O}\setminus\mathcal{A}_k$, since the set $\mathcal{S}$ itself is randomly chosen. Thus, we have:
\begin{align}
&\widehat{\Psi}(\mathcal{A}_k  \cup\{\alpha_{k+1}'\}) - \widehat{\Psi}(\mathcal{A}_k) \nonumber \\
&\geq  \Pr[\mathcal{S} \cap (\mathcal{O}\setminus \mathcal{A}_k ) \neq \emptyset]
\frac{\sum_{ o \in \mathcal{O}\setminus \mathcal{A}_k } ( \widehat{\Psi}( \mathcal{A}_k \cup\{o\}) - \widehat{\Psi}( \mathcal{A}_k ) ) }{|\mathcal{O}\setminus \mathcal{A}_k |} \nonumber \\
& \geq \frac{1-\epsilon}{B}  \sum_{ o \in \mathcal{O}\setminus\mathcal{A}_k } ( \widehat{\Psi}( \mathcal{A}_k \cup\{o\}) - \widehat{\Psi}( \mathcal{A}_k ) ) \nonumber \\
& \geq \frac{1-\epsilon}{B}  ( \widehat{\Psi}(\mathcal{O}) - \widehat{\Psi}( \mathcal{A}_k ) )
\label{eq:mu2}
\end{align}
where the second inequality is because of (\ref{eq:probS1}). The third inequality is due to the rule of diminishing returns. By combining (\ref{eq:mu1}) and (\ref{eq:mu2}) we obtain:
\begin{align}
\widehat{\Psi}(\mathcal{A}_{k+1}) - & \widehat{\Psi}(\mathcal{A}_k) =
\widehat{\Psi}(\mathcal{A}_k  \cup\{\alpha_{k+1}\}) - \widehat{\Psi}(\mathcal{A}_k) \nonumber \\
&\geq
\frac{(1-\epsilon) \frac{\sum_{t=1}^{\tau} \mu^t_{k+1}}{\tau} }{B}  ( \widehat{\Psi}(\mathcal{O}) - \widehat{\Psi}(\mathcal{A}_{k}) )
\end{align}

By induction, we can show that:
\begin{align}
\widehat{\Psi}(\mathcal{A}_{B}) &\geq  \Big( 1 - \big( 1 - \frac{(1-\epsilon) \frac{\sum_{k=1}^{B} \sum_{t=1}^{\tau} \mu^t_{k}}{B \tau} }{B}  \big)^{B} \Big) \widehat{\Psi}(\mathcal{O}) \nonumber \\
&\geq \big( 1 - e^{-(1-\epsilon) \frac{\sum_{k=1}^{B} \sum_{t=1}^{\tau} \mu^t_{k}}{B \tau} } \big) \widehat{\Psi}(\mathcal{O})
\end{align}
Since the $\mu^t_{k}$ values are drawn from a distribution with mean value $\mu$, it will be $\frac{\sum_{k=1}^{B} \sum_{t=1}^{\tau} \mu^t_{k}}{B \tau} = \mu $ in expectation, which concludes the proof.

\appendix[Proof of Theorem 4]
Let $\mu^1_1,\dots,\mu^{\tau}_{B}$ be the marginal gain ratios associated with the $B\tau$ try-outs of the Stochastic greedy algorithm.
 Since $\mu^t_k \in (0,1)$, $\forall t,k$ with mean value $\mu$, we can apply the Chernoff bound and obtain for each $\gamma \in (0,1)$:
\begin{align}
\Pr [ \frac{1}{B \tau} \sum_{k=1}^{B} \sum_{t=1}^{\tau} \mu^t_k < (1-\gamma) \mu   ] < e^{- \frac{\gamma \mu B \tau}{2} }
\end{align}
Therefore, with probability $1-e^{- \frac{\gamma \mu B \tau}{2} }$, the empirical mean value will be at least as much as $(1-\gamma) \mu $. With the same probability, the performance will be at least as much as $1-e^{-(1-\epsilon)(1-\gamma)\mu}$.

\end{document}